# An Accelerated LIF Neuronal Network Array for a Large Scale Mixed-Signal Neuromorphic Architecture

Syed Ahmed Aamir*, *Student Member, IEEE,* Yannik Stradmann*, Paul Müller, Christian Pehle, Andreas Hartel, Andreas Grübl, Johannes Schemmel, *Member, IEEE,* and Karlheinz Meier, *Member, IEEE*

*Abstract*—We present an array of leaky integrate-and-fire (LIF) neuron circuits designed for the second-generation BrainScaleS mixed-signal 65-nm CMOS neuromorphic hardware. The neuronal array is embedded in the analog network core of a scaled-down prototype HICANN-DLS chip. Designed as continuous-time circuits, the neurons are highly tunable and reconfigurable elements with accelerated dynamics. Each neuron integrates input current from a multitude of incoming synapses and evokes a digital spike event output. The circuit offers a wide tuning range for synaptic and membrane time constants, as well as for refractory periods to cover a number of computational models. We elucidate our design methodology, underlying circuit design, calibration and measurement results from individual sub-circuits across multiple dies. The circuit dynamics match with the behavior of the LIF mathematical model. We further demonstrate a winner-take-all network on the prototype chip as a typical element of cortical processing.

## I. INTRODUCTION

THE architecture of digital microprocessors is fundamentally different from that of the central nervous system. While the brain is a massively parallel structure of neurons interconnected through synapses [1], microprocessors are mostly based on a von Neumann architecture [2], [3] with logic gates as the elementary primitives. The human brain consumes only approximately 20 W [4], while its performance as a general-purpose problem solver is still unmatched by any computer algorithm.

Taking inspiration from this biological feat, neuromorphic architectures not only adopt a non-von Neumann architecture by collocating memory close to the computational element, but also introduce massive parallelism, high energy efficiency, reconfigurability, fault tolerance, and integrate computational models of neural elements using CMOS technologies [5]–[7]. They compute at biological timescales or at a specified speed-up ("acceleration") factor. In particular, analog implementations integrate bio-physically inspired neuron models as computational elements in order to capture the rich temporal dynamics of the neuronal membrane.

Since their emergence in the late eighties [5], neuromorphic circuits and systems have evolved with a variety of large-scale architectures, model implementations and technologies [4], [8]. For example, [9]–[14] describe systems that adhere to the initial idea of exploiting weak-inversion characteristics of CMOS circuits to emulate ion flow dynamics. More recently, with the advancements in cognitive computing driven in part by the speech and object recognition made possible by deep neural networks, major industrial players [15], [16] explore neuromorphic architectures and integrate digital phenomenological neuron models in ultra deep-submicron technologies — the most recent being Intel's 14 nm FinFET Loihi chip [16].

Within EU's framework for brain-inspired computing [17], [18], the SpiNNaker and BrainScaleS systems present large-scale neuromorphic platforms. While SpiNNaker [19] uses arrays of many-core interconnected ARM-based microprocessor nodes and implements a software-defined neuron model, our approach as described within the BrainScaleS hardware system [20], [21] is a wafer-scale implementation with an analog physical neuron model and accelerated-time operation. The system utilizes an entire post-processed CMOS wafer for large scale integration by interconnecting multiple identical on-wafer chips [21].

In the second-generation BrainScaleS architecture [22], [23], we present, for the first time, a mixed-signal neuromorphic computing core that integrates a custom Single Instruction Multiple Data (SIMD) processor with the Analog Network Core (ANC) (comprised of the neuron array and synapse matrix) in 65 nm CMOS. We have previously described the plasticity subsystem of the chip [23], [24]. Here we present the first silicon implementation and characterization of the neuronal array [25] based on the Leaky Integrate and Fire (LIF) neuron model. We demonstrate its high tunability, reconfigurability, reliable parameter mapping through calibration and the network operation on a scaled-down prototype chip. The neuron array is tested in tight integration with the synapses and dedicated analog bias storage [26], [27] providing voltage and current biases for tuning the neuron's behavior. The presented chip features a decreased neuronal and synaptic count (32

*Both authors contributed equally to this work.

Manuscript submitted August 29, 2017, revised January 20, 2018; April 4, 2018; May 14, 2018, accepted May 21, 2018. This work has received funding from the European Union Seventh Framework Programme ([FP7/2007-2013]) under grant agreement nos. 604102 (HBP), 269921 (BrainScaleS), the Horizon 2020 Framework Programme ([H2020/2014-2020]) under grant agreement no. 720270 (HBP) as well as from the Manfred Stärk Foundation.

All authors are with the Kirchhoff Institute for Physics, Heidelberg University, D-69120 Heidelberg, Germany.

Email: {aamir,yannik.stradmann}@kip.uni-heidelberg.de.

neurons interconnected with 32×32 synapses instead of 512 neurons and 256×512 synapses), but integrates all essential elements of the scaled-up version (see Sec. II). We demonstrate a winner-take-all network as an example application. The detailed circuit description of synapses, the analog bias storage and the plasticity mechanism of the chip is omitted in this work.

In the subsequent sections, we describe the chip architecture of the HICANN-DLS prototype, as well as the design pre-considerations (Sec. II & Sec. III). The design, measurement, calibration and performance results of the neuron as well as of the individual sub-circuits are described in Sec. IV. A demonstration of a winner-take-all network is detailed in Sec. V. We discuss and compare our results with other large-scale neuromorphic systems in Sec. VI before concluding the paper in Sec. VII.

## II. THE HICANN-DLS CHIP

The BrainScaleS wafer-scale system [21], [28] is built in the first phase with 180 nm CMOS HICANN neuromorphic chips, operated $10^4$–$10^5$ times faster than biological timescale. The second-generation hardware features the enhanced High Input Count Analog Neural Network with Digital Learning System (HICANN-DLS) chips as the fundamental building block [22]. HICANN-DLS is a 65 nm CMOS mixed-signal system on-chip solution with integrated analog and digital cores and operated one thousand times faster than biological real-time. The chip is under development and an enhanced scaled-up version of the current prototype will replace the HICANN chip in the second-generation wafer-scale platform.

A simplified architecture of the initial chip prototype connected to its off-chip measurement system is sketched in Fig. 1a. The left half shows the ANC arranged in a columnar architecture of the edge-connected neuron array, the synapse matrix as well as the analog on-chip Capacitive Memory (Capmem) cells for the storage of tunable neuron voltage and current biases. The right half shows the digital part comprising of the SIMD plasticity processor and digital control logic. In this prototype version, the ANC contains 32 columns, each containing 32 synapse rows relaying current pulse events to a single neuron per column. Each neuron is configured by 14 current and 4 voltage biases, all of which are individually tunable per neuron as well as a 15-bit digital configuration bus. One global voltage bias is common to all neurons and provided by an additional global Capmem column, see Fig. 1a.

The Capmem voltage and current cells have 10-bit resolution and provide the neuron with tunable biases of 15 nA to 1 µA (current cells) and 200 mV to 1.8 V (voltage cells). The average drift of the capacitive storage is less than 1 LSB/20 ms and the cells are periodically refreshed at a rate of 1–2 kHz to minimize its effect [27]. The columnar arrangement of neurons, synapses and dedicated tunable biases make the system a highly configurable non-von Neumann architecture. Data transfer between the digital backend and various components of ANC is carried out in the form of data packets via the On-chip Multi-master Non-bursting Interface Bus-fabric (OMNIBUS) [24] (blue connections in Fig. 1a). The packets encode synaptic addresses, synapse matrix's row enables and the digital configuration of neurons and the Capmem. During network operation, incoming events reach the synapses via OMNIBUS containing target synapse addresses and row-wise enables. If a packet's synapse address matches the target synapse address, the target enables a 6-bit current-mode Digital to Analog Converter (DAC) for a short duration, essentially relaying a current pulse event to the neuron on either of the two dedicated lines per neuron column (2×32 in total) — one for excitatory events, another for inhibitory events. The digital input code of the 6-bit DAC inside each synapse corresponds to the *synaptic weight*. This is further highlighted in the left half of Fig. 1b. The digital *output events* evoked by each neuron are routed off-chip to the FPGA via a Serializer/Deserializer (SerDes), from where they are re-routed back into the synapse array. To provide a debug interface, the entire neuron array is connected to the chip pads by two dedicated pins, labeled $I_\text{stim}$ and $V_\text{readOut}$ (described in Sec. IV-D).

The system performs hybrid learning using parallel analog processing in the synapses together with arbitrarily programmable learning rules in the custom SIMD plasticity processor [23]. The synapses feature a dedicated implementation of Spike-timing-dependent plasticity (STDP) [29], where they measure the temporal correlation of pre– and post-synaptic events. The post-synaptic events from all neurons are therefore routed to the synapses. To facilitate STDP, the temporal correlation of input/output events (pre-post or post-pre) is stored as analog voltages inside each synapse. These analog voltages are digitized by an Analog to Digital Converters (ADCs) before being read by the plasticity processor [23]. The processor can then alter the the synaptic weights (stored in SRAM) in accordance with the implemented rule, thereby performing online learning.

The processor is a 32-bit implementation based on power instruction-set architecture with a custom SIMD unit. It processes 128-bit wide vectors of either eight or sixteen elements using vector slices. The current implementation features a single slice that processes 16 columns in parallel and iterates twice for 32 columns. All synapse rows are processed sequentially. The processor can be used for implementing arbitrary learning rules that make use of the available observables and modify the structure and parameters of a running network. It can additionally be used to perform local calibration algorithms that can be executed in parallel on large multi-chip systems.

Within the described architecture, the total area of all ADC channels, the neurons and the number of vector slices in the processor scale linearly with the number of columns.

## III. DESIGN METHODOLOGY

When it comes to neuron design, every large scale system adheres to certain parameters governed mainly by the system specifications and target applications. The design phase of this accelerated neuron circuit in general took the following pre-considerations:

### A. Neuron Model

Our choice of neuron model comes from the trade-off between a design that encapsulates most of the biological computational power versus the silicon area and design complexity. This entails that the emulated neuron models should be rich enough to replicate most computational studies, as well as sufficiently simple to integrate enough neurons for small functional networks on a single die. This is a good compromise to envision large-scale systems, also endorsed by other architectures, e.g., [30]. Further, Integrate and Fire (I&F) models with refractoriness and adaptation are known to reproduce the biological traces with good accuracy [31], [32]. Our models of choice are therefore low dimensional threshold-based I&F models such as [33], [34]. Therefore, in this initial prototype, we implement the LIF model as a base computational element. The subthreshold response of the LIF model can be defined by Kirchhoff's current law for the conservation of charge:

$$C_\text{mem} \frac{dV_\text{mem}}{dt} = -g_\text{leak} \cdot (V_\text{mem} - V_\text{leak}) + I \qquad (1)$$

and if $V_\text{mem} \geq V_\text{thresh}$,

$$V_\text{mem} \rightarrow V_\text{reset} \qquad (2)$$

where $V_\text{mem}$ is the potential across the neuronal membrane, $C_\text{mem}$ is the membrane capacitance, $V_\text{reset}$ and $V_\text{thresh}$ are well defined reset and threshold potentials, $V_\text{leak}$ models the leak potential and $g_\text{leak}$ is the leak conductance. $I$ is the sum of external current ($I_\text{stim}$) and excitatory ($I_\text{syn,exc}$) and inhibitory ($I_\text{syn,inh}$) synaptic currents. The synaptic inputs integrating these currents are exponentially decaying current-based inputs, whose time course can be defined as:

$$I_\text{syn}(t) = \sum_i \sum_f w_i e^{-\left(\frac{t-t_i^f}{\tau_\text{syn}}\right)} \Theta(t - t_i^f) \qquad (3)$$

where $w_i$ is the weight of the synapse connecting a pre-synaptic neuron to a post-synaptic neuron, $t_i^f$ denotes $f$th spike at a synapse $i$, $\Theta(x)$ is the Heaviside step function and $\tau_\text{syn}$ is the synaptic time constant.

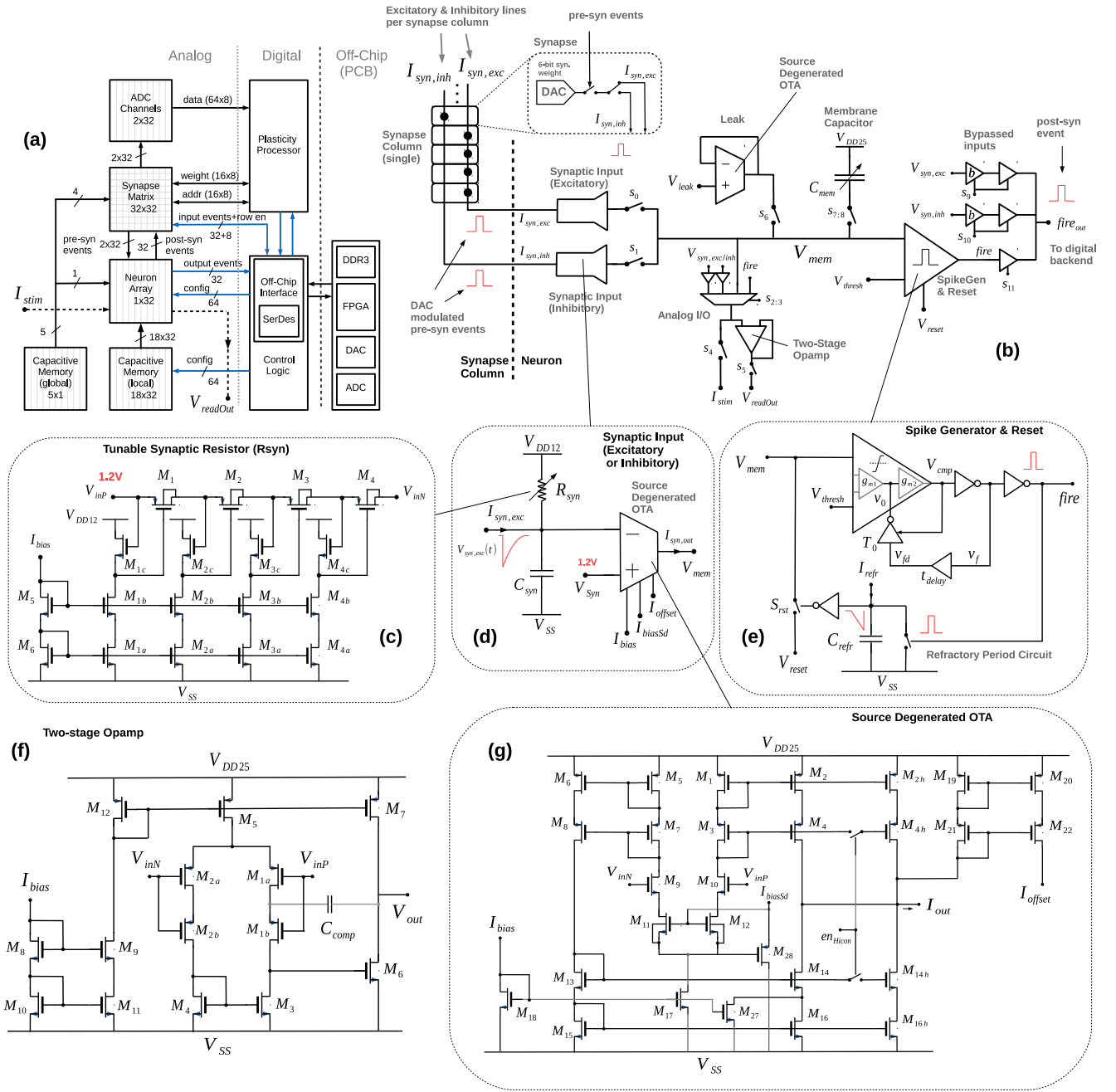

Fig. 1: (a) Architecture of the HICANN-DLS prototype chip and the measurement system. (b) The full circuit schematic of a single integrated neuron. (c) The schematic of the resistor used inside the synaptic input. (d) The architecture of the excitatory synaptic input (swapped terminals for inhibitory synaptic input). (e) The schematic of the spike pulse generator and reset circuit (implements refractory period duration). (f) The two-stage opamp schematic used inside the read-out buffer. (g) The schematic of the source-degenerated Operational Transconductance Amplifier (OTA) (used inside synaptic input and leak).

### B. Specifications and Parameter Range

A selected set of computational modeling studies [35]–[46] defines the tuning range and target specifications of the individual sub-circuits and eventually the neuron model parameters. Table I summarizes these identified ranges for membrane and synaptic time constants ($\tau_\text{mem}$, $\tau_\text{syn}$) as well as the refractory period ($\tau_\text{refr}$) in biological real-time. The presented hardware implementation uses current-based synaptic input. Because the reference studies employ conductance-based synapse models, an increased leak conductance by a factor of five is required to be able to emulate the reduction of the effective membrane time constant that is associated with high conductance states [47], [48].

| Parameter | Min. [ms] | Max. [ms] |
|---|---|---|
| $\tau_\text{mem}$ | 7 | 50 |
| $\tau_\text{syn}$ | 1 | 100 |
| $\tau_\text{refr}$ | 0 | 10 |

TABLE I: Target specifications of the neuron model parameters.

### C. Debug and Testability

The neuron circuit provides the possibility to debug individual sub-circuits or disconnect individual terms for verification and calibration. For example, a digitally configurable bypass mode is featured that evokes a single output spike per input synaptic event. The aforementioned digitally switchable terms are realized by the transmission gate switches $S_n$ in Fig. 1b. An integrated voltage buffer reads out the membrane potential as well as the activity on both shared synaptic input lines within each neuron circuit.

### D. Power and Area

Power reduction for large-scale integration is made possible at two levels: First, the Capmem in this chip prototype can provide bias currents as low as 15 nA. This saves significant power consumption, since the circuits shift to moderate or weak inversion. Further, the unused sub-circuits are power-gated. The sub-circuit design has been optimized for area/power by utilizing MOS gate capacitors, usage of thin and thick-oxide transistors and dual voltage supply options, wherever possible. As a result, the single neuron circuit consumes approximately 10 µW power (14.4 µW if both synaptic inputs are used).

### E. Calibration

A major goal of the presented design was the possibility to calibrate individual sub-circuits as well as the entire neuron. The circuit has therefore been verified pre-silicon with focus on mismatch compensation using the available Monte Carlo models. Post-silicon, all 32 neurons have been calibrated for mismatch across multiple dies, the respective results are shown in Sec. IV. Utilizing these datasets, we can provide a mapping between LIF model quantities and circuit-level parameters by using fractional polynomial fits for the individual sub-circuits.

## IV. CIRCUIT DESIGN AND MEASUREMENTS

The neuron schematic designed for the current prototype chip is shown in Fig. 1b connected to a single column of the synapse matrix. Starting from the left, it shows a single synapse column from which two output lines emanate — one carrying excitatory pre-synaptic events and another inhibitory pre-synaptic events. Each pre-synaptic event enables a 6-bit DAC for a configurable duration of 10 ns–320 ns, which outputs a current pulse on either of the two synaptic lines. At the neuron side, these current pulses are received by the synaptic input circuits (one excitatory, one inhibitory). These circuits model the synaptic dynamics with an exponential kernel inside each neuron circuit (Fig. 1d). The latter integrate current from both synaptic inputs onto a capacitor $C_\text{mem}$ that models the neuronal membrane. The membrane can discharge itself through a separate *Leak* circuit towards the resting potential $V_\text{leak}$. Once the voltage reaches a threshold $V_\text{thresh}$, the circuit *SpikeGen* triggers a digital pulse event (see *fire* in Fig. 1b,e) and brings the membrane back to a reset potential $V_\text{reset}$ via a *Reset* module. This circuit also adds the respective refractory period, essentially by clamping the membrane to $V_\text{reset}$ for a duration governed by the time it takes to toggle the inverter again. An *Analog IO* block reads out debug voltages and injects (test) stimulation current onto the membrane. The neuron schematic also highlights the digitally controlled interconnecting switches labeled $S_{0-11}$. These are realized as transmission gates and are meant to test, debug and switch-off individual terms whenever needed.

Each of the incoming input events to either excitatory or inhibitory synaptic inputs can be made to directly trigger output events (see *fire*$_{\text{out}}$ in Fig. 1b), bypassing the current integration on the neuron membrane. Inverters with shifted trip-points driven directly by the voltage drop on the input synaptic lines make this possible — they are enabled by digital inputs $S_{9,10}$ and tri-state inverters at the output. This "bypass mode" is useful for testing certain modules of the system (e.g. event routing) without relying on configured neurons. It is disabled during nominal neuron operation.

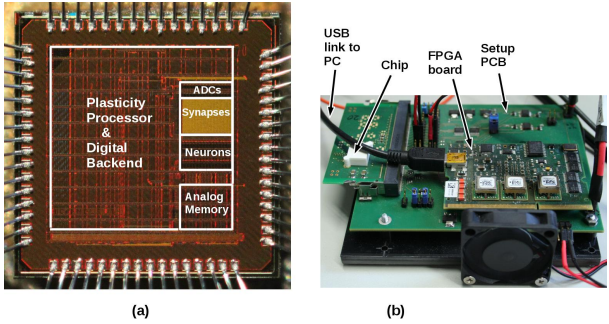

Fig. 2: (a) Chip micrograph. (b) Prototype measurement system.

The integrated array of 32 neuron circuits occupies 200 μm × 376 μm of die area. Each neuron in turn occupies 200 μm × 11.76 μm. The described prototype chip has an area of 1.9 × 1.9 mm$^2$ and is fabricated in a low-K 1P9M 65-nm low-power digital CMOS process. The bonded die and the measurement setup are shown in Fig. 2. All measured analog data presented within this work has been acquired using a Keithley 2635B Sourcemeter for static current measurements and a LeCroy Wavesurfer 44Xs digital oscilloscope for dynamic signals. The chip measurement framework comprises of an on-board Xilinx Spartan-6 FPGA system that takes command packets via a USB interface. The chip and all neuron parameters are directly programmable via a C++/Python based software environment. The neuron block specifications are summarized in Table VII. Before demonstrating the neuron operation, we describe the design and measurements of each individual sub-circuit.

### A. Synaptic Input

The synaptic input circuit provides the exponential synaptic dynamics to incoming events inside each neuron circuit (Eq. III-A). The circuit integrates short current pulses (synaptic events) arriving from 32 synapse circuits in a single column onto an RC integrator before converting them into an equivalent current with the help of a linear transconductor. The architecture is shown in Fig. 1d. The current pulse events from 32 synapse circuits are received on the shared line labeled $I_{\text{syn,exc}}$ in Fig. 1b and shown as an input terminal in Fig. 1d. Each incoming input pulse event drops the voltage on this line (nominally at 1.2 V), which is then recovered with the synaptic time constant $\tau_{\text{syn}}$. This voltage drop is proportional to the strength of the incoming pulse event. The synaptic time constant is varied by the tunable resistor $R_{\text{syn}}$, while the integration capacitor $C_{\text{syn}}$ is fixed at 1 pF. In the current prototype, this capacitance is mostly contributed by a metal capacitor, but it will eventually be realized from the line parasitics alone — as the number of input synapses will increase in the scaled-up final chip. The second OTA input $V_{\text{syn}}$ is kept at approximately 1.2 V, the precise value can be altered for calibration purposes. The realized amplifiers can cancel the effect of input offset voltage at its output using a tunable bias current $I_{\text{biasOff}}$.

*1) Transconductor:* The designed transconductor is a source-degenerated OTA architecture, whose output current has a linear dependence on input differential voltage within a limited range, such that $I_{\text{out}} = G_{\text{m}}(V_{\text{in+}} - V_{\text{in-}})$. Where $G_{\text{m}}$ is the OTA transconductance. The OTA's main bias current (labeled $I_{\text{bias}}$) helps to vary this transconductance, whereas a second bias (labeled $I_{\text{biasSd}}$) adjusts the proper biasing point of the source degeneration pair. The OTA architecture is shown in Fig. 1g. Transistors $M_1$-$M_4$, and $M_5$-$M_8$ form cascode current mirror loads, whereas $M_{9,10}$ form the input pair. Transistors $M_{11,12}$ are a source degeneration pair, acting therefore as degenerating resistors to lower the gain. Their bias points are tunable externally using a separate bias $I_{\text{biasSd}}$.

Note the presence of two separate stages at the output: First, the output stage current multiplier formed by transistors $M_{2h,4h}$ and $M_{14h,16h}$ — it provides extra output current in parallel to the existing output stage formed by $M_{2,4}$ and $M_{14,16}$. A digital switch labeled $en_{\text{hiCon}}$ enables this current multiplication. This is useful when realizing a higher conductance — for example in the leak term to ensure short membrane time constants. The leak term (see Sec. IV-C) also uses this OTA and the output current multiplier is implemented only there. Secondly, another parallel circuit — a current mirror formed by $M_{19,20}$ and $M_{21,22}$ is used as an offset calibration circuit at the OTA output. $I_{\text{offset}}$ is then the calibration current that is set equal to the residual offset current, caused for example as a result of input offset voltage between the OTA terminals. The current mirror steals this current from the output stage of the OTA that directly sends current to the membrane.

*2) Synaptic Resistor:* The synaptic resistor $R_{\text{syn}}$ is a tunable resistor designed using bulk-drain connected devices [49]–[51] and shown in Fig. 1c. The architecture of the resistor uses a series of four pmos bulk-drain connected devices (labeled $M_1$ to $M_4$) between the two terminals $V_{\text{inP}}$ and $V_{\text{inN}}$. Each pmos device connects its bulk terminal to its drain instead of the nominal bulk configuration. They provide a linear increase in drain current with increasing source-drain voltage. Series stacked devices are used to reduce the source-drain potential well below the threshold voltage. The total voltage drop across the two resistor terminals can be as large as a couple of hundred millivolts for the case of larger incoming events. For example, a 10 μA event (if 4 ns long), can drop the synaptic input line $I_{\text{syn,exc/inh}}$ to 1 V, creating a drop of 200 mV across the terminals. The resistor therefore provides almost linear operation within this range. The pull-up nature of this synaptic input resistor ensures that source potentials are always higher than the drains. This prevents the two terminals from getting swapped and possibly consume higher currents, since it is then a nominal pmos configuration (bulk connected to source terminal). A single bias current $I_\text{bias}$ tunes the gate voltage of all subsequent bulk-drain connected devices via cascode current mirrors, labeled $M_{5,6}$ and $M_{Xa}$–$M_{Xb}$, where $X$ denotes the respective devices of each pmos stage. The devices marked $M_{Xc}$ help tune the transistor bias points. Table II summarizes the transistor dimensions for the designed resistor. Notice that in order to provide larger resistance the channel lengths of the bulk-drain connected devices $M_{1-4}$ are relatively long. We have measured synaptic input circuits on up to three different dies. The characteristic curves of the resistor within the desired voltage range are shown in Fig. 4. The figure plots the current vs. the potential difference by varying the resistor's bias ($I_\text{bias}$ in Fig. 1c). The equivalent swept $I_\text{bias}$ and the resulting resistances determined from a linear fit are shown. Note that at very low bias currents the resistance is very large (bottom-blue curve) and the resistance values are inclined for an exponential increase as a function of bias current. Notice the presence of finite voltage offsets, as the plotted traces cut the zero-nA line at a potential difference of about 15 mV across the terminals. This is due to supply drop and can be corrected for by tuning the input $V_\text{syn}$ (second OTA terminal) during operation.

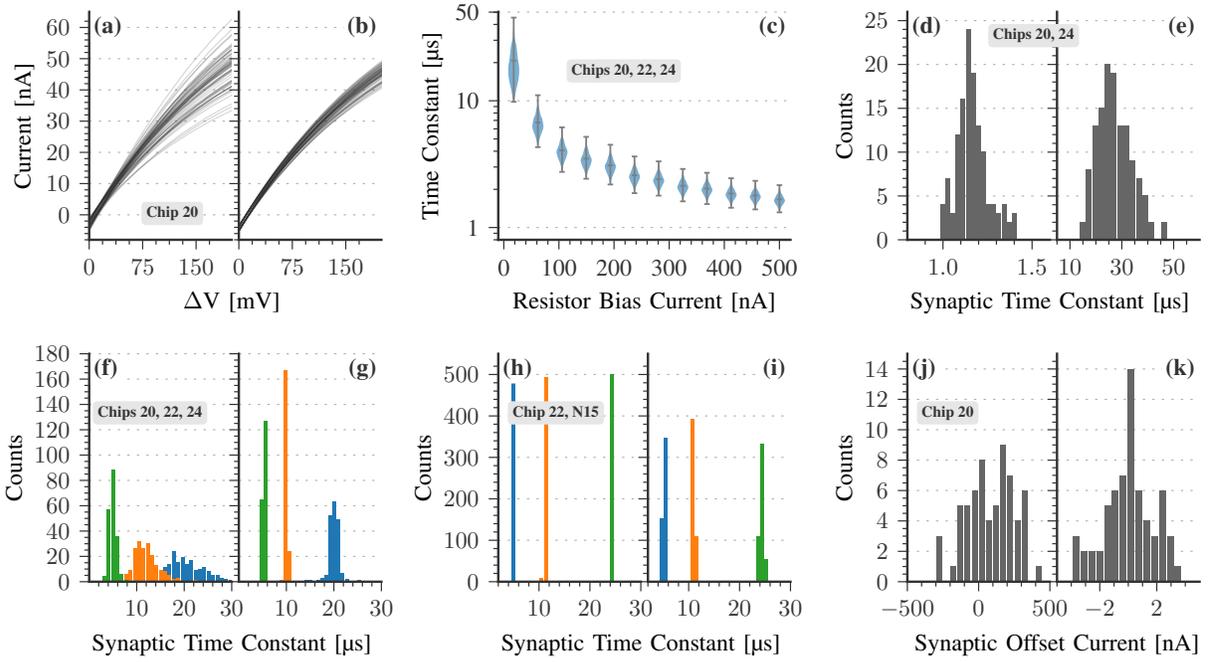

Fig. 3: (a) Variation of the characteristic curves for the 64 synaptic input resistors of a single chip, with bias set to 80 nA (pre-calibrated). (b) Post-calibration resistor curves with time constant of all synaptic inputs set to equal the mean of those in (a). (c) Tuning the synaptic time constants of 192 synaptic input circuits by varying the resistor bias current (d,e) Distribution of the minimum and maximum range of the achieved synaptic time constants. (f) Pre-calibration distribution of synaptic time constants with a mean of 5 µs, 10 µs and 20 µs. (g) Post calibration: the target value of (f) is processed through individual polynomial fits. (h) Trial-to-trial variation of synaptic time constants for programmed values of 5 µs, 10 µs and 25 µs. (i) Variation in synaptic time constants measured from a train of 500 continuous input events. (j) Pre-calibrated synaptic output offset current with $V_\text{syn}$ at 1.2 V for all OTAs. (k) Residual synaptic current after individually adjusting $V_\text{syn}$ and $I_\text{biasOff}$.

| Device | Width/Length [µm/µm] |
|---|---|
| $M_{1,2,3,4}$ | 0.15/1.0 |
| $M_{1c,2c,3c,4c}$ | 0.15/5.0 |
| $M_{1b,2b,3b,4b}$ | 0.15/0.5 |
| $M_{1a,2a,3a,4a}$ | 0.15/0.5 |
| $M_{5,6}$ | 0.75/0.5 |

TABLE II: The device dimensions of the bulk-drain connected synaptic resistor.

We further show our measurement results from all 64 resistors (both synaptic inputs), configured for the same mid-range resistance of about 4 MΩ in the desired voltage range in Fig. 3a. The traces show triode curves for typical MOS devices, since the bulk-drain connected devices are in fact biased in linear region to adjust the resistive range. The variation in traces are due to device mismatch, largely contributed from the biasing stages ($M_{Xa,Xb,Xc}$) that help set the bias points of each individual device. This is minimized in Fig. 3b after applying the calibration described in Sec. IV-A4.

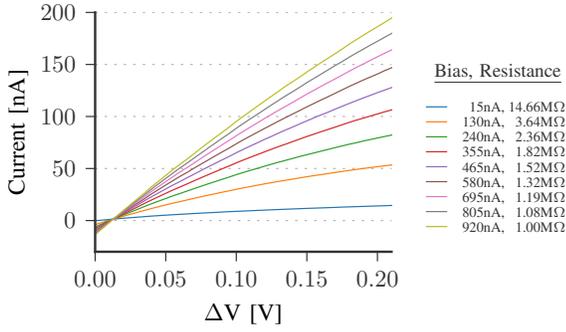

Fig. 4: Tunability of the synaptic resistor. The current flowing through $R_{syn}$ is measured as a function of the voltage drop across the resistor terminals, as the resistor bias is swept.

*3) Full Circuit:* Finally, the measured results demonstrating the range of possible synaptic time constants $\tau_{syn}$ are shown in Fig. 3c. The data has been acquired by measuring three separate dies, each having 64 synaptic input circuits. The plot shows a range of available time constants as well as their mean value as the bias of the tunable resistor is swept from 0 to 0.5 µA. The plot shows a non-linear tuning curve which steepens substantially in the low current regime. It corresponds to the tuning behavior of synaptic resistor shown in Fig. 4.

The variation in different achievable time constants, hence is larger when we tune longer synaptic time constants, compared to shorter ones. This is also evident from the histogram plot of Fig. 3d,e, where a distribution of both the shortest and longest synaptic time constants (bias settings of 1 µA and 15 nA) are plotted for 128 samples. Given these distributions, the range of available synaptic time constants is between 1.24 µs and 20.5 µs for one-sigma single sided quantiles (see Table III).

*4) Calibration:* The input-referred offset of the synaptic input OTA, if not canceled, can make the neuron membrane integrate unwanted output offset current. To calibrate this, we use the OTA reference voltage $V_{syn}$ as well as the offset compensating bias $I_{biasOff}$, both of which are locally tunable parameters. Using $V_{syn}$ we cancel the output offset to a first approximation. The residual offset is trimmed by sweeping $I_{biasOff}$ and searching the root of a linear fit. Fig. 3j,k shows the measurement of all 64 synaptic inputs' output offset currents on a single die pre- and post-calibration. The post-calibration residual current is below 5 nA for all samples, which corresponds to approximately 20 mV deviation in the membrane resting potential at a membrane time constant of 10 µs. The curves shown in Fig. 3c are fitted with second order fractional polynomials, which reduces the relative spread of individual time constants. Fig. 3f,g shows the spread of three different synaptic time constants pre- and post-calibration from three different dies. The standard deviation for these datasets is depicted in Tab. IV. Finally, we demonstrate the trial-to-trial variation for three different time constants within a single neuron. In Fig. 3h, the mean time constant of 500 incoming synaptic events is measured after reprogramming from a different time constant. Most samples are clearly restricted to the same bin. Fig. 3i on the other hand shows the synaptic time constant variation of 500 incoming events for the aforementioned three configurations.

### B. Spike Generator and Reset

This circuit is responsible for evoking a digital output event as an indication of spike occurrence, once the membrane reaches a specified voltage threshold $V_{thresh}$. It also resets the membrane voltage to a fixed reset potential $V_{reset}$ and adds a refractory period $\tau_{refr}$ before the membrane starts integrating again. The circuit consists of a comparator whose output stage is reset after a finite delay via a feedback loop, as shown in Fig 1e. Once the membrane $V_{mem}$ reaches the threshold $V_{thresh}$, the comparator outputs a logic 'high' ($V_{OH}$). After a finite time delay $t_{delay}$, the comparator output stage is pulled down to logic 'low' ($V_{OL}$). This essentially generates a digital voltage pulse signal *fire*. The *fire* pulse marks the spike occurrence and is routed as a single event through to the digital backend. Furthermore, it also clamps the membrane $V_{mem}$ to $V_{reset}$ by toggling the switch $S_{rst}$ (see Fig 1e). This happens as the voltage over the capacitor $C_{refr}$ (which constantly integrates a tunable bias current $I_{refr}$) is reset by the incoming pulse. This initiates the neuron's refractory period ($\tau_{refr}$) that lasts until the voltage across the capacitor $C_{refr}$ triggers the inverter again to disconnect the switch $S_{rst}$ (Fig 1e). This marks the end of refractory period and the membrane $V_{mem}$ is then free to integrate again.

The delay-cell designed for the spike pulse generator (labeled $t_{delay}$) is a tunable current-starved delay element [52] typically set to 100 ns. The comparator is a simple two-stage architecture similar to an uncompensated two-stage op-amp that provides high gain to drive outputs to the desired output levels ($V_{OH}$ or $V_{OL}$). To realize long refractory periods without a large $C_{refr}$, the bias current $I_{refr}$ from the respective Capmem cell is divided 10 times (fixed), reducing the effective bias currents to 1.5 nA (min.) and 100 nA (max.).

Our measurements of two dies (32 circuits each) suggest that the refractory period can be tuned successfully over a range from 1.11 µs up to about 137 µs, as shown in Fig. 5a,b ($1\sigma$ quantile, see also Table III). The variation in longer time constants is relatively large due to very small input bias current (1.5 nA) being integrated on $C_{refr}$. The circuit has been designed for much longer refractory periods than what is known through biology to cater for artificial models such as [53].

*1) Calibration:* Since the duration of the refractory period is solely determined by the current $I_{refr}$ (Fig. 1e), this variation is catered for by measuring $\tau_{refr}$ as a function of $I_{refr}$ and applying a second order fractional polynomial fit for each neuron. In Fig. 5c,d, the distributions of three different refractory periods are shown pre- and post-calibration. Notice the larger spread towards smaller bias currents and therefore higher time constants, which is significantly reduced after calibration (c.f. Tab. IV). Fig. 5e shows the calibrated membrane potential of all 32 neurons on a single die after spike occurrence. All neurons are configured to a reset potential of 200 mV, a spike threshold of 1.2 V, a resting potential of 1 V and — in order to produce steep, comparable edges — to a short membrane time constant of 250 ns. The figure shows two traces per neuron, one programmed for a refractory time of 10 µs and the other for 25 µs. Note how all neurons reach the resting potential with little remaining spread in the time constant. The visible variations of the resting potential are due to residual offset currents from the synaptic inputs (c.f. Sec. IV-A4), as well at least 31 mV of input offsets of the leak OTAs (estimated through simulation using Monte Carlo device models).

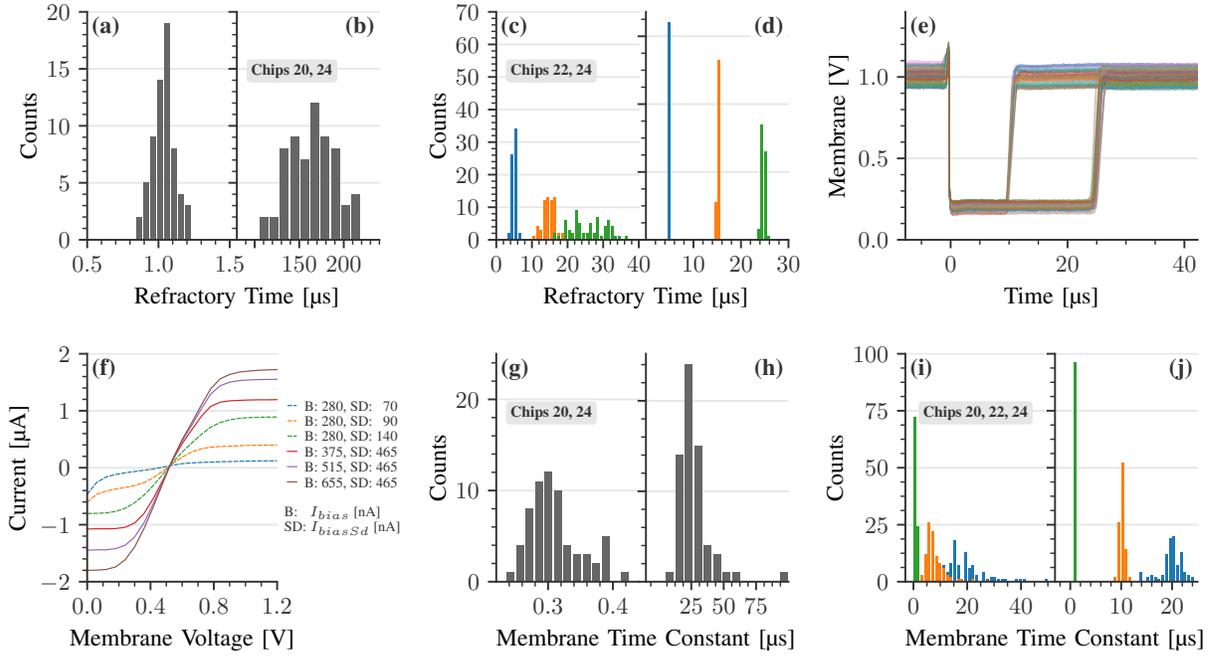

Fig. 5: (a,b) The distribution of minimum and maximum achievable refractory time periods. (c,d) Pre- and post-calibration results of refractory period circuits, set for a mean of 5 µs, 15 µs and 25 µs respectively. (e) Calibrated membrane traces showing a post-spike response of different on-chip neurons with refractory times of 10 µs and 25 µs. (f) Output current of the leak term as the membrane potential is swept with the resting potential fixed at 0.52 V. (g,h) Distribution of the minimum and maximum achievable membrane time constants. (i,j) Pre- and post-calibration distribution for $\tau_\mathrm{mem}$ of 1 µs, 10 µs and 20 µs.

### C. Membrane Leak and Capacitance

The membrane conductance has been realized using a linear transconductor whose output terminal is fed back to its negative input as shown in Fig. 1b. In this configuration, the total conductance across the input and output terminals equals the transconductance $G_\mathrm{m}$. The usage of linear transconductors to realize conductance is already well known in neuromorphic architectures [5], its architecture is described in Sec. IV-A1.

A two-bit configurable gate oxide capacitor provides a total of 2.36 pF membrane capacitance. This is implemented in multiples of 590 fF for layout symmetry, switchable between 2.36 pF, 1.77 pF or 590 fF via $S_{7-8}$ (see Fig. 1b). To minimize distortions introduced by the inherently non-linear behavior of MOS capacitors, the respective transistors were chosen to be thick-oxide devices. Within the membrane's full dynamic range of up to 1.2 V, the bulk-gate drop can therefore be kept above 1.3 V and the device in inversion region where the gate capacitance is constant.

| Parameter [1] | min ($1\sigma$, $3\sigma$)[µs] | max ($1\sigma$, $3\sigma$)[µs] |
|---|---|---|
| $\tau_\mathrm{refr}$ | 1.11, 1.24 | 137.5, 104.5 |
| $\tau_\mathrm{syn}$ | 1.24, 1.41 | 20.5, 13.6 |
| $\tau_\mathrm{mem}$ [2] | 0.35, 0.41 | 16.6, 11.3 |

[1] Single-sided $1\sigma$ ($3\sigma$) quantiles of 84.13% (99.86%)
[2] Measured using $C_\mathrm{mem}$ = 2.36 pF; The min./max. $\tau_\mathrm{mem}$ ($1\sigma$, $3\sigma$) for $C_\mathrm{mem}$ = 570 fF is estimated to be 0.08, 0.10 µs and 4.11, 2.82 µs respectively

TABLE III: Measured range of time constants from data acquired across two dies.

In Fig. 5g,h, we show the distribution of achieved membrane time constants with the largest selectable membrane capacitance. The distribution highlights the minimum and maximum achievable time constants ($\tau_\mathrm{mem}$) from a total of 64 leak circuits on two different dies. It is shown that the neuron can set time constants, as small as 0.35 µs and up to 16.6 µs (one-sigma quantiles) and can be further reduced by selecting different capacitor configurations (c.f. Table III).

Fig. 1b shows how the leak term uses the transconductor in a feedback configuration, as opposed to the synaptic input where the OTA is in open loop. In feedback configuration, the OTA transconductance $G_\mathrm{m}$ is the overall conductance of the leak term, defining membrane time constant as $\tau_\mathrm{mem} = C_\mathrm{mem}/G_\mathrm{m}$.

|  | target | pre-calibration | post-calibration |
|---|---|---|---|
| $\tau_{\text{refr}}$ | 5 µs | 0.50 µs (9.95%) | 0.14 µs (2.83%) |
|  | 15 µs | 1.84 µs (12.3%) | 0.12 µs (0.77%) |
|  | 25 µs | 4.88 µs (18.9%) | 0.51 µs (2.06%) |
| $\tau_{\text{syn}}$ | 5 µs | 0.75 µs (15.2%) | 0.14 µs (2.63%) |
|  | 10 µs | 2.49 µs (20.7%) | 0.18 µs (1.75%) |
|  | 20 µs | 4.01 µs (19.9%) | 1.12 µs (5.56%) |
| $\tau_{\text{mem}}$ | 1 µs | 0.19 µs (19.3%) | 0.03 µs (3.03%) |
|  | 10 µs | 3.14 µs (40.9%) | 0.46 µs (4.53%) |
|  | 20 µs | 8.34 µs (42.3%) | 2.27 µs (11.5%) |

TABLE IV: Pre- and post calibration standard deviations for the histograms shown in Fig. 3 and Fig. 5.

Fig. 5f shows six measured traces plotting the output current of the leak term as a function of the membrane voltage. Three of them sweep the main bias current $I_{\text{bias}}$ keeping $I_{\text{biasSd}}$ constant, while the remaining three sweep $I_{\text{biasSd}}$, while $I_{\text{bias}}$ is held constant. The traces highlight how lowering the source degeneration bias linearizes the response and allows for a wider range of membrane voltages. This is because the degeneration resistors, implemented as transistors $M_{11}, M_{12}$, are shifted from linear region to the saturation region of the weak inversion regime, where their on-resistance $R_s$ is as large as few MΩ at low $I_{\text{biasSd}}$. This effectively linearizes the curve as the OTA's transconductance $G_m$ approximates to $1/R_s$.

*1) Calibration:* The membrane time constant can be tuned by three parameters: A two-bit variable membrane capacitor and the two biases ($I_{\text{bias}}$, $I_{\text{biasSd}}$) of the leak OTA. Despite the low gain curves and consequently wide range for lower bias values of $I_{\text{biasSd}}$ (Fig. 5f), tuning is done using the former two parameters. The circuit achieves membrane time constants between 0.35 µs and 16.6 µs within a linear range of about 230 mV. The reason for not relying on $I_{\text{biasSd}}$ is twofolds — first, low values of $I_{\text{biasSd}}$ contribute a large input offset, causing an output residual current, that is not calibrated for by tuning $V_{\text{leak}}$. Secondly, at lower values of $I_{\text{biasSd}}$, overall transconductance $G_m$ contributed by the OTA is a non-monotonic function of $I_{\text{bias}}$. To avoid both effects, we calibrate the membrane time constant by adjusting the OTA's bias current ($I_{\text{bias}}$) and therefore trade linear range for robustness. We measure the relation between $\tau_{\text{mem}}$ and $I_{\text{bias}}$ at fixed $I_{\text{biasSd}}$ of 1 µA for a voltage decay from 600 mV towards a resting potential of 400 mV. The resulting curves are fitted by second degree fractional polynomials. Pre- and post-calibration variations of $\tau_{\text{mem}}$ are depicted in Fig. 5i,j for targets of 1 µs, 10 µs and 20 µs, the related standard deviations are shown in Tab. IV. It may be noted that not all neurons can achieve the maximum time time constant of 20 µs. These neurons are visible as outliers to the left of the depicted distribution.

### D. Analog Input/Output

We can inject constant current into the membrane, hold it to a fixed potential or read out debug voltages via the analog I/O sub-circuit (see $I_{\text{stim}}$ and $V_{\text{readOut}}$ in Fig. 1b).

| Output range | 0.1–2.1 V |
|---|---|
| Load | ≥ 16 pF ‖ 10 MΩ[a] |
| –1-dB bandwidth | 1.15 MHz[b] |
| Gain | 62.6 dB[c] |
| Power | 127 µW[c] |
| Input offset | 14 mV |
| Slew rate | 2 V/µs |
| Compensation cap. | 600 fF |

TABLE V: Measured specifications of the read-out buffer.

[a]Estimated off-chip load
[b]Measured after the LPF formed by output Tx-gate and parasitic load
[c]Simulation only

| Device | Width/Length [µm/µm] |
|---|---|
| $M_{1a,1b,2a,2b}$ | 2.4/0.28 |
| $M_{3,4}$ | 0.8/0.38 |
| $M_6, M_7$ | 4.0/0.38 & 4.0/0.39 |
| $M_{12}, M_5$ | 0.15/0.75 & 1.6/0.39 |
| $M_{9,11}$ | 3.6/0.38 |
| $M_{8,10}$ | 0.6/0.38 |

TABLE VI: The device sizes of the read-out amplifier.

The monitored output voltage is multiplexed from among the membrane potential or incoming synaptic activity at any of the two synaptic input lines. The Opamp designed for the buffer is a two-stage amplifier with split-length indirect compensation [54], [55] and is shown in Fig. 1f. The buffer has a –1-dB bandwidth of 1.15 MHz when driving large off-chip loads at a maximum power consumption of 127 µW. The close-loop opamp bandwidth is filtered by the output low-pass filter formed by transmission gate and the off-chip load. In the current design, each neuron buffer directly drives the shared read-out line that terminates at the output pad. In comparison to Miller compensation, the use of indirect compensation ensures smaller area, higher bandwidth, and reduced power consumption [55]–[57]. During nominal neuron operation, the amplifier is switched off, the stated power is only consumed during analog read-out (therefore not reported in Sec. III-D). The measured amplifier specifications are summarized in Table V and the transistor dimensions used in this design are listed in Table VI.

### E. Full Circuit Characterization

Here we describe the measurements and performance of the complete neuron circuit. Fig. 6 shows equidistant incoming events at the excitatory synaptic input line (bottom trace) and the resulting membrane trace (top). The reset potential is set to 0.2 V, the leak potential to 0.58 V and the spiking threshold to 1.2 V. The input synaptic lines (bottom trace) have a constant 1.2 V potential unless an input event arrives. At event arrival, this level drops and is recovered with a time constant $\tau_{\text{syn}}$. Each incoming event increases the membrane potential, eventually

| | |
|---|---|
| Neuron model | Leaky I&F |
| No. of neurons | 32 |
| Area of single neuron | 11.76 × 200 μm² |
| Voltage supply | 2.5/1.2 V |
| Process | 65-nm CMOS |
| Speed-up (acceleration) factor | ×1000 |
| Global parameters | 1 voltage bias [1] |
| Local (individual) parameters | 18 (14 I-bias, 4 V-bias)[1] |
| Configurability | 15-bit digital bus |
| Membrane capacitor (max.) | 2.36 pF (2-bit config.) |
| Input synaptic event (max.) | 10 μA, 4 ns pulses[2] |

[1] available from on-chip tunable capacitive memory cells
[2] amplitude and length of each current pulse emitted by the synapse circuit

TABLE VII: A summary of neuron array specifications.

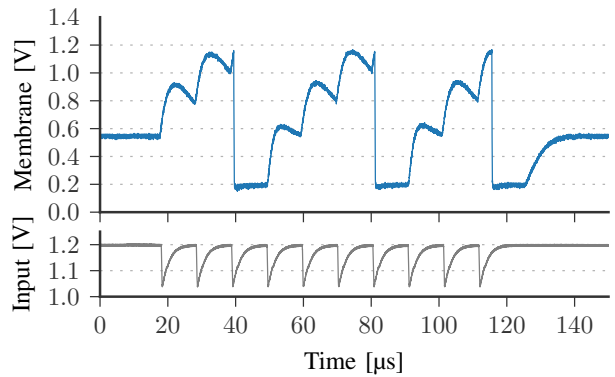

Fig. 6: Measured results — *bottom*: synaptic input potential indicating incoming events as evaluated on $I_{\text{syn,exc}}$ line, *top*: resulting membrane response to the synaptic events.

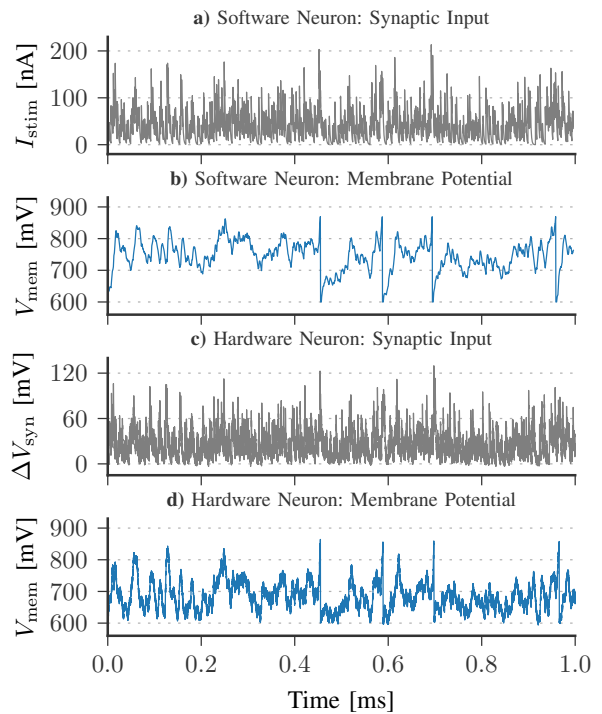

Fig. 7: A one-to-one comparison of the emulated neuron vs. software simulation.

evoking a spike, followed by a membrane reset to initiate the refractory period duration.

We further demonstrate the neuron parameter set from a typical modeling study [43] and compare the response properties of our emulated on-chip neuron, with that of a simulated accelerated LIF model neuron (Eq. III-A). We use the Brian spiking neural network simulator [58] to simulate the ideal LIF neuron. The on-chip neuron circuit here is uncalibrated and tuned manually such that the most pronounced non-ideal effects are canceled for. This includes the output residual current of synaptic input due to input offset voltage between the OTA terminals. Further, time-constants are fine-tuned and synaptic weights are adjusted to match the spike times of the software simulation. The synaptic and membrane time constants ($\tau_{\text{syn}}$, $\tau_{\text{mem}}$) are set as 1.5 μs and 10 μs, with a refractory period $\tau_{\text{refr}}$ of 2 μs (×1000 speed-up from biological real-time). The threshold is set to 870 mV and the leak and reset potentials are 600 mV. A random spike train stimulus containing $32 \times 20$ events (each synapse relaying 20 pre-synaptic events) is injected into both neurons and their corresponding membrane traces are plotted in Fig. 7. Fig. 7a shows the input current as a result of random input stimulus of the software neuron, while Fig. 7b shows its corresponding membrane voltage. In the hardware domain, we monitor the on-chip voltage on the excitatory synaptic input line, as its voltage drop (defined as $\Delta V_{\text{syn}} := V_{\text{Syn}} - V_{\text{syn,exc}}$, see Fig. 1d) is in proportion to the charge of each incoming synaptic event. A larger voltage drop corresponds here to higher input synaptic current (Fig. 7c). Fig. 7d shows the measured membrane voltage of the on-chip neuron. Comparing the dynamics of both membrane traces, one finds the RMS of their common-mode corrected difference to be 45 mV. We see that, when presented with the same input, the on-chip neuron has similar spike-times and the membrane voltage follows the time course of the software simulation.

### F. Power Consumption

In order to estimate the chip prototype's power consumption, the total current drawn by the chip during operation of the network described in Sec. V has been measured. From this measurement, we infer a mean full-chip power consumption of $(48.62 \pm 0.04)\,\text{mW}$. This figure includes prototype-specific consumers — especially the 2.5 V full-swing IO channels that communicate each spike off-chip to the controlling FPGA. This IO power is expected to scale sublinearly with increasing chip size.

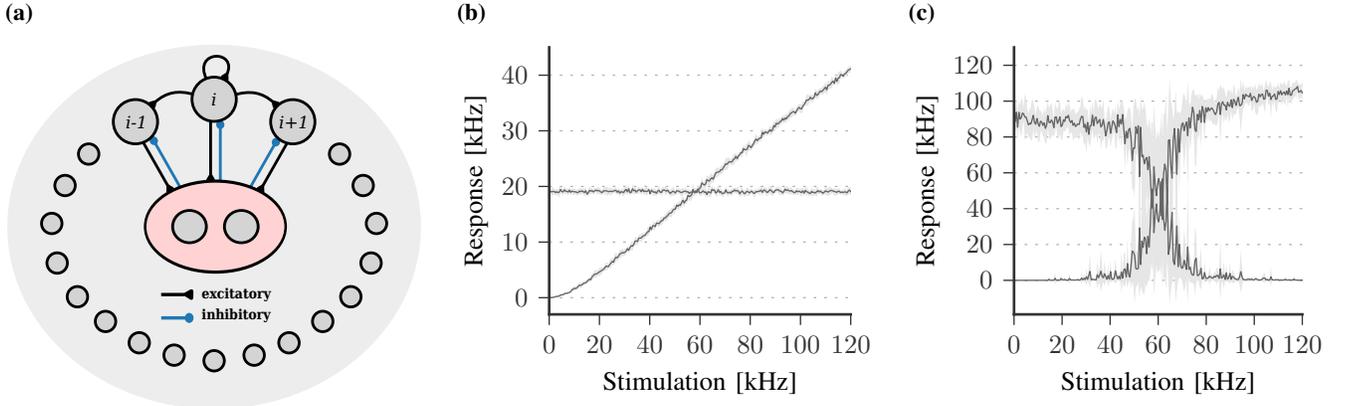

Fig. 8: (a) Schematic diagram of the implemented network topology. All neurons in the network form self- and direct-neighbor excitatory projections, as well as to central inhibitory neurons. Both inhibitory neurons in turn inhibit the ring neurons and do not make connections between themselves. (b) Network response for two competing input stimuli $r_1$ and $r_2$ before the network routing has been established: One neuron receives Poisson noise with an average rate of $r_1 = 50\,\text{kHz}$, another neuron receives Poisson noise with variable frequency between $r_1 = 0\,\text{kHz}$ and $r_1 = 120\,\text{kHz}$. The mean firing rate of these neurons and their next neighbors is depicted as an average over 10 trials with different random seeds for the input stimulus. (c) Network response for two competing input stimuli $r_1$ and $r_2$ after network routing has been established: The Poisson input is as specified in Fig. 8b.

An estimation of the mean energy cost of a single spike has been compiled by comparing the aforementioned result against the power consumption of an idle chip (without stimulating spikes) with the same network configuration. Due to the greatly reduced number of expensive outgoing traffic when the system is idle, the resulting number has to be understood as an upper bound for the on-chip spike cost. We find $E_{\text{spike}} \leq (790 \pm 170)\,\text{pJ}$ for the presented network.

## V. A Winner-Take-All Network

We demonstrate a soft winner-take-all circuit [59]–[61] using the on-chip LIF neuronal array. From an array of 32 neurons, 20 are used to form a ring of excitatory neurons, each having self-excitation, nearest-neighbor excitatory projection, and excitatory projections to two common inhibitory neurons. The inhibitory pool of two neurons in turn inhibit all neurons in the ring and do not make any projection between themselves. The resulting network topology is shown in Fig. 8a. All neurons are chosen randomly and are calibrated as described in the previous sections. To verify the network's behavior, two opposite (equidistant) neurons in the ring network are stimulated with Poisson stimuli of various frequencies between 0 kHz and 120 kHz. Fig. 8c shows the average firing rates of both competing sub-populations. A sub-population here is defined as the stimulated neuron and its two direct neighbors. The stimulus to one is fixed at 50 kHz, while to the other is being swept. The swept stimulus has an increment of 0.5 kHz between separate runs, accounting to 240 different stimuli runs within the specified range. Each experiment has been run for 100 ms. The line/highlighted region in Fig. 8c shows the average and standard deviation over 10 independent runs; each with a different random seed for the Poisson input. Due to the speed-up of the presented neuron array (compared to biological time scales), the total network emulation time of four minutes (10 trials, each consisting of 240 different stimuli à 100 ms) corresponds to 2.8 days of biological network activity. The mean firing rate for both populations during 10 control runs with disabled spike connectivity (routing) is shown in Fig. 8b. As in Fig. 8c, the mean firing frequency of one single sub-population is plotted — as there are no connections, both neighbors are quiet and population firing rate is scaled roughly by a factor of three.

## VI. Discussion

We have presented an array of LIF neurons designed for the initial prototype of our HICANN-DLS chip. The paper summarized comprehensive measurement and calibration results to demonstrate the usability of the realized neuron array. It has been demonstrated that the statistical spread of the time constants can be minimized significantly, once they are calibrated for device mismatch and corner variations. The transconductor architecture features an offset cancellation mechanism that reduces the residual synaptic output offset by two orders of magnitude. The usability of the circuit has been demonstrated by its low trial-to-trial variation. In comparison to synaptic time constants, the membrane time constants show wider spread, since the former was tuned via the synaptic resistor which has a better control compared to leak OTA biases. We conclude that both synaptic as well as membrane time constants need improvement to cover longer durations in a future revision. The circuit evokes a digital event as an output spike that also initiates the refractory period — a very broad duration has been demonstrated with a statistical spread that is widely reduced after calibration. The presented prototype is ready to run local calibration routines using the embedded processor, thereby enabling highly parallel calibration of multiple neurons in large-scale systems. Such local calibration algorithms are subject to future development.

|  | TrueNorth | Neurogrid | This Work |
|---|---|---|---|
| CMOS tech. [nm] | 28 | 180 | 65 |
| Architecture | digital | analog subth. | analog |
| Speed-up | ×1 | ×1 | ×1000 |
| Neuron Model | augmented LIF | quad. I&F | LIF |
| Biophysical Dynamics[a] | No | Yes | Yes |
| Neuron Area [$\mu m^2$] | 2900[b] | 1800 | 2352[c] |
| Local Parameters | Yes | No | Yes |
| Parameter Area [$\mu m^2$] | – | – | 2688[d] |

[a] tunable $\tau_{refr}$, $\tau_{mem}$, $\tau_{syn}$
[b] multiplexed 256 times per time step
[c] including excitatory and inhibitory synaptic inputs
[d] area of 18 analog on-chip biases per neuron for current work; not comparable to TrueNorth's 124-bit SRAM-based parameter config. or Neurogrid's shared biases

TABLE VIII: An overview of neuron model specifications in large-scale neuromorphic architectures.

|  | TrueNorth | Neurogrid | Hi-DLS |
|---|---|---|---|
| Die Area [$mm^2$] | 430 | 168 | 32 |
| Neurons/core [#] | 4k | 64k | 512 |
| Neurons/time[a][#] | 1M | 64k | 512k |
| Synapses/core [#] | 64k | 256k[b] | 131k |
| SynOps/sec [GSOPS] | 58 | – | 16k |

[a] emulated neurons per timescale: HICANN-DLS accelerates ×1000; TrueNorth multiplexes ×256
[b] each synapse population (4× per neuron) counted as individual synapse

TABLE IX: An overview of single-chip characteristics of large-scale neuromorphic systems with scaled-up HICANN-DLS chip. Data plotted in Fig. 9.

To verify the characteristics of the emulated physical neuron, a comparison with a software-simulated ideal LIF neuron is made: the number of spikes as well as the membrane dynamics are found to be in close correspondence. Finally, using the chip routing resources, the FPGA and system software stack, a small soft winner-take-all network is realized. A basic functionality of the circuit is shown by feeding Poisson input to two competing sub-populations. It is pertinent to note that usage of accelerated hardware reduced the network run-time (compared to biological real-times) by three orders of magnitude, enabling a platform feasible to study long-term developmental processes in the nervous system.

We further compare our chip's neuron implementation with two state-of-the-art neuromorphic implementations, namely the IBM's TrueNorth system [15] and Stanford University's Neurogrid architecture [9]. The TrueNorth neuron is a phenomenological LIF model that implements various synthetic, arithmetic and logical operations along with real-time spiking behaviors. Being a phenomenological model, the temporal dynamics of the synaptic and membrane time constants ($\tau_{syn}$, $\tau_{mem}$) are not tunable implementations. Despite a 28 nm CMOS implementation, the digital circuit occupies more area (2900 $\mu m^2$), reduced effectively by the possibility of time-multiplexing. Alternate phenomenological implementations such as the one presented by Loihi's 14 nm architecture offer slightly worse neuron density [16]. Compared to this, Neurogrid — like BrainScaleS — offers biologically plausible dynamics and tunable time constants. However, all 64 k neurons in a single *Neurocore* share their parameters, whereas the current work has dedicated neuron parameters. Neurogrid implements subthreshold MOS dynamics, which typically allow for compact neurons due to small currents leading to less capacitor area for given time constants. Both Neurogrid and TrueNorth operate in real-time, as opposed to our accelerated approach. The neuron area occupied here is slightly larger than Neurogrid neuron — in which about 500 $\mu m^2$ is occupied by membrane capacitor (MOScap). The neuron model we implement in the current prototype is a LIF model, compared to Neurogrid's two compartment quadratic model. The presented modular neuron architecture has been enhanced to an Adaptive-Exponential I&F (AdEx) model with multiple compartments in a recently submitted chip prototype [62]–[65]. Table VIII summarizes this comparison against the TrueNorth and Neurogrid neurons.

The measured energy per output spike (Sec. IV-F) includes the off-chip routing and has been reduced by providing local (on-die) routing in the next prototype. We have not measured energy per synaptic activation in our chip which has been measured at 941 pJ and 26 pJ for implemented networks in Neurogrid and TrueNorth systems. We expect that the upper bound of $(790 \pm 170)$ pJ of energy/spike will reduce with more neurons and input synapses in the scaled-up system. The energy consumed in the stated neuromorphic architectures are orders of magnitude lower than, for example, a quad-core Xeon CPU [66] simulating a spiking neural network and still more efficient compared to the network emulated on many-core embedded systems, e.g. on the SpiNNaker system [67].

Fig. 9 presents a multi-dimensional overview of this prototype chip with the scaled-up HICANN-DLS as well as the single-chip TrueNorth and Neurogrid implementations. The chip area, number of physical neurons and synapses are shown against peak synaptic operations per second (SOPS) and emulated neurons in a given time-scale. The resulting pentagons reflect the architectural preference of each implementation. Data about peak SOPS was not available for Neurogrid. The TrueNorth chip with the largest die area (430 $mm^2$) time-muliplexes the 4096 physical neuron instances to realize the maximum number of neurons and synapses, while Neurogrid operates in real-time and features the largest physical neuron count (65 k per *Neurocore*). The HICANN-DLS chip features a 6-bit STDP enabled synapse circuit, which cannot be compared with Neurogrid's analog-diffusor network and synapse-population circuits (which we count here as individual synapse) — and neither with TrueNorth's 1-bit crossbar synapses. For the least die area, HICANN-DLS maximizes synaptic operations per second due to the time-acceleration and possibility of high input data rates. We emphasize that Fig. 9 and its corresponding data given in Table IX may infer a very limited set of architectural and operational features integrated by each large-scale system. It highlights how the scaled-up chip of current prototype will feature against other single-chip building blocks. A direct comparison pertaining to varying architectural designs and system capabilities is not possible.

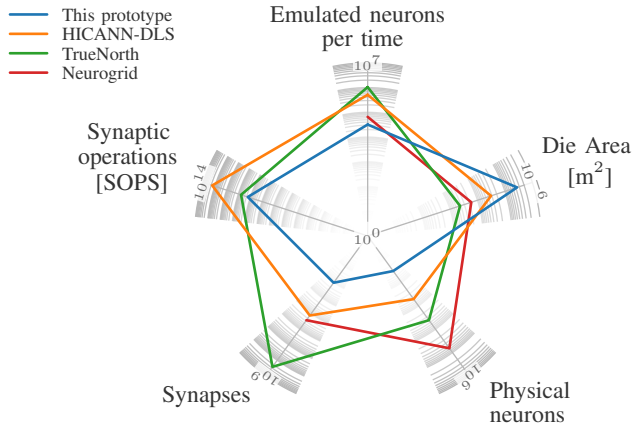

Fig. 9: Overview of the single-chip implementations in contemporary large-scale neuromorphic architectures.

## VII. Conclusion

We have presented the design, measurement, calibration and application of our mixed-signal 65-nm CMOS LIF neuronal array for the second-generation BrainScaleS platform. The neurons cover a wide tunable parameter range to act as a general-purpose element for many computational models. Having demonstrated a functional scaled-down prototype of the 65-nm chip in tight-integration with synaptic matrix and analog capacitive memory, we march on towards the development of an enhanced scaled-up HICANN-DLS chip for our large-scale bio-physically inspired neuromorphic system.

## Author Contribution and Acknowledgment

S.A.A. wrote the manuscript, designed the neuron circuits and implemented the on-chip array. Y.S. performed chip measurements, post tape-out calibration and the network demonstration. P.M. devised the pre tape-out calibration and provided biological benchmarks. C.P. implemented the external FPGA spike routing. A.H. and A.G. implemented the digital backend and performed top-level verification. J.S. was the overall system architect and conceived the initial OTA and resistor architectures, later implemented by S.A.A. The conceptual advice was given by K.M.

The authors would like to extend their gratitude to Matthias Hock for support during design and measurement.

## References

[1] E. R. Kandel, J. H. Schwartz, and T. M. Jessell, *Principles of Neural Science*, 4th ed. New York: McGraw-Hill, 2000.

[2] J. von Neumann, "First draft of a report on the edvac," *IEEE Annals of the History of Computing*, vol. 15, no. 4, pp. 27–75, 1993.

[3] J. von Neumann, *The computer and the brain*. New Haven, CT, USA: Yale University Press, 1958.

[4] K. Meier, "Special report : Can we copy the brain? - the brain as computer," *IEEE Spectrum*, vol. 54, no. 6, pp. 28–33, June 2017.

[5] C. A. Mead, *Analog VLSI and Neural Systems*. Reading, MA: Addison Wesley, 1989.

[6] G. E. Moore, "Cramming more components onto integrated circuits," *Proceedings of the IEEE*, vol. 86, no. 1, pp. 82–85, 1998.

[7] G. Indiveri and S.-C. Liu, "Memory and information processing in neuromorphic systems," *Proceedings of the IEEE*, vol. 103, no. 8, pp. 1379–1397, 2015.

[8] S. Furber, "Large-scale neuromorphic computing systems," *Journal of Neural Engineering*, vol. 13, no. 5, p. 051001, 2016.

[9] B. V. Benjamin *et al.*, "Neurogrid: A mixed-analog-digital multichip system for large-scale neural simulations," *Proceedings of the IEEE*, vol. 102, no. 5, pp. 699–716, 2014.

[10] R. J. Vogelstein *et al.*, "Dynamically reconfigurable silicon array of spiking neurons with conductance-based synapses," *IEEE Transactions on Neural Networks*, vol. 18, no. 1, pp. 253–265, Jan 2007.

[11] Y. Wang and S.-C. Liu, "A two-dimensional configurable active silicon dendritic neuron array," *IEEE Transactions on Circuits and Systems I: Regular Papers*, vol. 58, no. 9, pp. 2159–2171, 2011.

[12] S. Brink, S. Nease, P. Hasler, S. Ramakrishnan, R. Wunderlich, A. Basu, and B. Degnan, "A learning-enabled neuron array IC based upon transistor channel models of biological phenomena," *IEEE Transactions on Biomedical Circuits and Systems*, vol. 7, no. 1, pp. 71–81, 2013.

[13] N. Qiao *et al.*, "A reconfigurable on-line learning spiking neuromorphic processor comprising 256 neurons and 128k synapses," *Frontiers in Neuroscience*, vol. 9, p. 141, 2015.

[14] E. Chicca, F. Stefanini, C. Bartolozzi, and G. Indiveri, "Neuromorphic electronic circuits for building autonomous cognitive systems," *Proceedings of the IEEE*, vol. 102, no. 9, pp. 1367–1388, Sept 2014.

[15] P. A. Merolla *et al.*, "A million spiking-neuron integrated circuit with a scalable communication network and interface," *Science*, vol. 345, no. 6197, pp. 668–673, 2014.

[16] M. Davies *et al.*, "Loihi: A neuromorphic manycore processor with on-chip learning," *IEEE Micro*, vol. 38, no. 1, pp. 82–99, January 2018.

[17] H. Markram, "The human brain project," *Scientific American*, vol. 306, no. 6, pp. 50–55, 2012.

[18] FACETS, "Fast Analog Computing with Emergent Transient States – project website," http://www.facets-project.org, 2010.

[19] S. B. Furber *et al.*, "The SpiNNaker project," *Proceedings of the IEEE*, vol. 102, no. 5, pp. 652–665, May 2014.

[20] K. Meier, "A mixed-signal universal neuromorphic computing system," in *Proc. IEDM*, Dec 2015, pp. 4.6.1–4.6.4.

[21] J. Schemmel, D. Brüderle, A. Grübl, M. Hock, K. Meier, and S. Millner, "A wafer-scale neuromorphic hardware system for large-scale neural modeling," in *Proc. ISCAS*, 2010, pp. 1947–1950.

[22] J. Schemmel, "BrainScales 2: A novel architecture for analog accelerated neuromophic computing and hybrid plasticity," *Internal document, ASIC Lab., Kirchhoff-Institute for Physics*, 2017.

[23] S. Friedmann, J. Schemmel *et al.*, "Demonstrating hybrid learning in a flexible neuromorphic hardware system," *IEEE Trans. BioCAS*, vol. 11, no. 1, pp. 128–142, Feb 2017.

[24] S. Friedmann, "A new approach to learning in neuromorphic hardware," Ph.D. dissertation, Ruprecht-Karls-Universität Heidelberg, 2013.

[25] S. A. Aamir, P. Müller, A. Hartel, J. Schemmel, and K. Meier, "A highly tunable 65-nm CMOS LIF neuron for a large scale neuromorphic system," in *Proc. ESSCIRC*, Sept 2016, pp. 71–74.

[26] M. Hock, "Modern semiconductor technologies for neuromorphic hardware," Ph.D. dissertation, Ruprecht-Karls-Universität Heidelberg, 2014.

[27] M. Hock, A. Hartel, J. Schemmel, and K. Meier, "An analog dynamic memory array for neuromorphic hardware," in *Proc. ECCTD*, Sep. 2013, pp. 1–4.

[28] C. Bartolozzi *et al.*, *Neuromorphic Systems*. John Wiley and Sons, Inc., 2016.

[29] H. Markram, J. Lübke, M. Frotscher, and B. Sakmann, "Regulation of synaptic efficacy by coincidence of postsynaptic aps." *Science*, vol. 275, pp. 213–215, 1997.

[30] J. V. Arthur and K. A. Boahen, "Silicon-neuron design: A dynamical systems approach," *IEEE Transactions on Circuits and Systems I: Regular Papers*, vol. 58, no. 5, pp. 1034–1043, 2011.

[31] R. Jolivet *et al.*, "Integrate-and-fire models with adaptation are good enough: predicting spike times under random current injection," *NIPS*, vol. 18, pp. 595–602, 2006.

[32] W. Gerstner, W. Kistler, R. Naud, and L. Paninski, *Neuronal Dynamics*. Cambridge University Press, 2014.

[33] L. Lapicque, "Recherches quantitatives sur l'excitation electrique des nerfs traitee comme une polarization," *Journal de Physiologie et Pathologie General*, vol. 9, pp. 620–635, 1907.

[34] R. Brette and W. Gerstner, "Adaptive exponential integrate-and-fire model as an effective description of neuronal activity," *J. Neurophysiol.*, vol. 94, pp. 3637 – 3642, 2005.

[35] A. Destexhe, "Self-sustained asynchronous irregular states and Up/Down states in thalamic, cortical and thalamocortical networks of nonlinear integrate-and-fire neurons." *Journal of Computational Neuroscience*, vol. 3, pp. 493 – 506, 2009.

[36] A. Destexhe and D. Contreras, "Neuronal computations with stochastic network states," *Science*, vol. 314, no. 5796, pp. 85–90, 2006.

[37] B. Nessler *et al.*, "Bayesian computation emerges in generic cortical microcircuits through spike-timing-dependent plasticity," *PLoS Computational Biology*, vol. 9, no. 4, p. e1003037, 2013.

[38] T. P. Vogels and L. F. Abbott, "Signal propagation and logic gating in networks of integrate-and-fire neurons," *J Neurosci*, vol. 25, no. 46, pp. 10 786–95, Nov 2005.

[39] G. Deco and V. K. Jirsa, "Ongoing cortical activity at rest: criticality, multistability, and ghost attractors," *The Journal of Neuroscience*, vol. 32, no. 10, pp. 3366–3375, 2012.

[40] R. Naud, N. Marcille, C. Clopath, and W. Gerstner, "Firing patterns in the adaptive exponential integrate-and-fire model," *Biological Cybernetics*, vol. 99, no. 4, pp. 335–347, Nov 2008.

[41] M. A. Petrovici, B. Vogginger, P. Müller, O. Breitwieser *et al.*, "Characterization and compensation of network-level anomalies in mixed-signal neuromorphic modeling platforms," *PloS one*, vol. 9, no. 10, 2014.

[42] M. A. Petrovici, I. Bytschok, J. Bill, J. Schemmel, and K. Meier, "The high-conductance state enables neural sampling in networks of LIF neurons," *BMC Neuroscience*, vol. 16, no. Suppl 1, p. O2, 2015.

[43] J. Kremkow, L. Perrinet, G. Masson, and A. Aertsen, "Functional consequences of correlated excitatory and inhibitory conductances in cortical networks." *J Comput Neurosci*, vol. 28, pp. 579–594, 2010.

[44] M. Pospischil, Z. Piwkowska, M. Rudolph, T. Bal, and A. Destexhe, "Calculating event-triggered average synaptic conductances from the membrane potential," *J. Neurophysiology*, vol. 97, p. 2544, 2007.

[45] T. Masquelier and G. Deco, "Network bursting dynamics in excitatory cortical neuron cultures results from the combination of different adaptive mechanism," *PloS one*, vol. 8, no. 10, p. e75824, 2013.

[46] S. Millner, "Development of a multi-compartment neuron model emulation," Ph.D. dissertation, Heidelberg University, 2012.

[47] D. Pare *et al.*, "Impact of spontaneous synaptic activity on the resting properties of cat neocortical pyramidal neurons in vivo," *J Neurophysiol*, vol. 79, no. 3, pp. 1450–60, 1998.

[48] A. Destexhe *et al.*, "The high-conductance state of neocortical neurons in vivo," *Nature Rev. Neurosci.*, vol. 4, pp. 739–751, 2003.

[49] F. Cannillo *et al.*, "Bulk-drain connected load for subthreshold MOS current-mode logic," *Electron. Lett.*, vol. 43, no. 12, June 2007.

[50] F. Cannillo *et al.*, "Nanopower subthreshold MCML in submicrometer CMOS technology," *IEEE Transactions on Circuits and Systems I: Regular Papers*, vol. 56, no. 8, pp. 1598–1611, Aug 2009.

[51] A. Tajalli, E. J. Brauer, Y. Leblebici, and E. Vittoz, "Subthreshold source-coupled logic circuits for ultra-low-power applications," *IEEE J. Solid-State Circuits*, vol. 43, no. 7, pp. 1699–1710, July 2008.

[52] D. K. Jeong, G. Borriello, D. A. Hodges, and R. H. Katz, "Design of PLL-based clock generation circuits," *IEEE J. Solid-State Circuits*, vol. 22, no. 2, pp. 255–261, Apr 1987.

[53] M. Petrovici, J. Bill, I. Bytschok, J. Schemmel, and K. Meier, "Stochastic inference with spiking neurons in the high-conductance state," *Physical Review E*, vol. 94, no. 4, October 2016.

[54] V. Saxena and R. J. Baker, "Indirect compensation techniques for three-stage fully-differential op-amps," in *Proc. MWCAS*, Aug 2010, pp. 588–591.

[55] S. A. Aamir, P. Harikumar, and J. J. Wikner, "Frequency compensation of high-speed, low-voltage CMOS multistage amplifiers," in *International Symposium on Circuits and Systems*, May 2013, pp. 381–384.

[56] V. Saxena and R. J. Baker, "Compensation of CMOS op-amps using split-length transistors," in *Proc. MWSCAS*. IEEE, 2008, pp. 109–112.

[57] S. A. Aamir, P. Angelov, and J. J. Wikner, "1.2-v analog interface for a 300-msps HD video digitizer in core 65-nm CMOS," *IEEE Transactions on VLSI Systems*, vol. 22, no. 4, pp. 888–898, April 2014.

[58] D. Goodman and R. Brette, "Brian: a simulator for spiking neural networks in Python," *Front. Neuroinform.*, vol. 2, no. 5, 2008.

[59] J. Lazzaro, S. Ryckebusch, M. Mahowald, and C. Mead, "Winner-take-all networks of o (n) complexity," in *NIPS*, vol. 1, 1988, pp. 703–711.

[60] R. Hahnloser *et al.*, "Digital selection and analogue amplification coexist in a cortex-inspired silicon circuit," *Nature*, vol. 405, no. 6789, pp. 947–951, 2000.

[61] T. Pfeil *et al.*, "Six networks on a universal neuromorphic computing substrate," *Frontiers in Neuroscience*, vol. 7, p. 11, 2013.

[62] S. A. Aamir, "Mixed-signal circuit implementation of spiking neuron models," Ph.D. dissertation, Karlsruhe Institute of Technology, 2018.

[63] S. A. Aamir, P. Müller, L. Kriener, G. Kiene, J. Schemmel, and K. Meier, "From LIF to AdEx neuron models: accelerated analog 65 nm CMOS implementation," in *Proc. BioCAS*, Oct. 2017, pp. 1–4.

[64] S. A. Aamir, P. Müller, G. Kiene, L. Kriener, Y. Stradmann, A. Grübl, J. Schemmel, and K. Meier, "A mixed-signal structured AdEx neuron for accelerated neuromorphic cores," *IEEE Transactions on Biomedical Circuits and Systems*, 2018.

[65] J. Schemmel, L. Kriener, P. Müller, and K. Meier, "An accelerated analog neuromorphic hardware system emulating NMDA- and calcium-based non-linear dendrites," in *Proc. IJCNN*, May 2017, pp. 2217–2226.

[66] P. S. Paolucci *et al.*, "Power, energy and speed of embedded and server multi-cores applied to distributed simulation of spiking neural networks: ARM in NVIDIA Tegra vs Intel Xeon quad-cores," *CoRR*, vol. abs/1505.03015, 2015.

[67] E. Stromatias *et al.*, "Scalable energy-efficient, low-latency implementations of trained spiking deep belief networks on SpiNNaker," in *Proc. IJCNN*, July 2015, pp. 1–8.